\documentclass[aps,pre,superscriptaddress,twocolumn,showpacs,floatfix]{revtex4}
\usepackage{graphics,dcolumn,amsmath,amssymb}
\usepackage{color}
\usepackage{epsfig}
\begin{document}
                                                        
\title{Networking genetic regulation and neural computation: Directed network topology and its effect on the dynamics}

\author{Andreas  \surname{Gr\"{o}nlund}}
\email{gronlund@tp.umu.se}
\affiliation{Department of Physics, Ume{\aa} University,
901 87 Ume{\aa}, Sweden}

\begin{abstract}
Two different types of directed networks are investigated, transcriptional regulation networks and neural networks.
The directed network structure are studied and also shown to reflect the different processes taking place on the networks. 
The distribution of influence, identified as the the number of downstream vertices, are used as a tool for investigating random vertex removal.
In the transcriptional regulation networks we observe that only a small number of vertices have a large influence.
The small influences of most vertices limit the effect of a random removal to in most cases only a small fraction of vertices in the network.
The neural network has a rather different topology with respect to the influence, which are large for most vertices.
To further investigate the effect of vertex removal we simulate the biological processes taking place on the networks. Opposed to the presumpted large effect of random vertex removal in the neural network, the high density of edges in conjunction with the dynamics used makes the change in the state of the system to be highly localized around the removed vertex.
\end{abstract}


\maketitle

\section{Introduction}
In recent years complex networks have drawn a great attention from the physics community.
Various measures have been introduced in order to capture the function and form of specific networks. An observed feature that many networks show are a scale free, or at least wide distribution of the vertex degree, which is given a popular and well cited explanation in \cite{ba:model}. 
Other studies includes measures of clustering, assortative mixing \cite{mejn:assmix}, betweenness centrality \cite{free:cen,bart:betw,mejn:rwbetw}.
For a review of the recent work on networks see \cite{mejn:rev,born:book,doromen:rev}.
Many of the networks appearing in the real world are directed and naturally the structure of two networks can be fundamentally different when the direction of the edges are considered, even if the overall structure might be alike when the direction of the edges are not considered.
Two examples of real world networks that are naturally directed are neural networks and transcriptional regulation networks.
The former is the network of neurons where neurons are connected in a directed fashion where the axons of each neuron connects to one of another neuron's dendrites and in this way building up a directed network in which signals are sent (axons) and received (dendrites) by the individual neurons.
In the transcriptional regulation networks, the vertices represents proteins and the edges are representing one proteins transcriptional regulation (positive and/or negative) of another protein. 
The cause of regulation is the attachment of a regulator protein to an operator position located on the DNA upstream of the gene coding for the regulated protein or if more than one protein, operon.
The attachment responses either in an up-regulation or down-regulation of the transcription rate by RNAp of the specific gene and thus the production of the protein.
The reasons for regulation are many of which one example is energy saving in poor nutritional environment since synthesis of RNA and protein both are energy expensive processes.
Another example is the regulation of enzymes.
The best studied case of enzyme induction involves the enzymes of lactose degradation in \emph{Escherichia\ coli}.
Only in the presence of lactose the enzymes that are necessary to utilize lactose as a carbon and energy source are synthesized.
But it is not just a matter of the presence of lactose.
If both glucose and lactose is present \emph{E.\ coli} chooses glucose. This is transcriptionally regulated via both positive and negative control.

The networks used in this paper are the neural network of the nematode \emph{Caenorhabditis\ elegance} (NNCE) \cite{white:420}, the transcriptional regulaton network of the bacteria \emph{E.\ coli} (TREC) \cite{alon:ecoli}  and the transcriptional regulaton network of yeast, \emph{Saccharomyces\ cerevisiae} (TRSC) \cite{Lee:yeast}.

\section{Structural properties}
\begin{figure}[htb]
\begin{center}
\includegraphics[width=0.95\columnwidth]{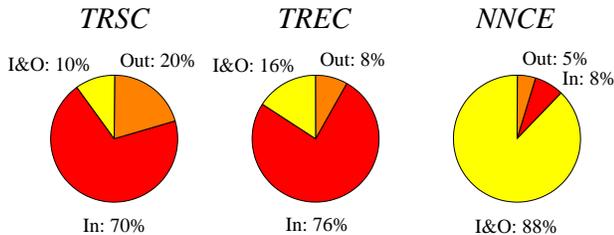}
\caption{The distribution of vertices with only incoming arcs, only outgoing arcs, and with both incoming and outgoing arcs.}
\label{Fig1}
\end{center}
\end{figure}

A directed edge is in the literature formally termed arc, which also will be the term used in this paper.
Because of the direction of the arcs one is able to follow directed paths in the network, representing the flow of information, the chain of command, or some other flow in the network.
Depending on the system ``living on'' the network, the structure might look very different when the direction of the arcs in the different networks are taken into account.
To get a first picture of what is going on in the networks we look at the distribution of the number of vertices with just outgoing arcs, only incoming arcs, and with both outgoing and incoming arcs.
In a network in which information is flowing like the neural network a significant fraction of vertices should have both incoming and outgoing arcs in order to transport information between different parts of the network.
Figure \ref{Fig1} shows the distribution of the three different types of vertices in the networks.
The neural network consists of mostly \emph{interneurons}, that is neurons with both incoming and outgoing arcs, and have a low fraction of \emph{sensory neurons} (only outgoing arcs) and \emph{motor neurons} (only incoming arcs). For more information on interneurons, sensory neurons and motor neurons see \cite{dale:neur}.
The regulation networks have a different structure where the number of vertices of both incoming and outgoing arcs are suppressed and the network is dominated by vertices of only incoming links.

Since the different neurons play a different role and have different functions, a natural question to ask would be if this is reflected in the degree of the different vertices.
In the neural network the sensory neurons receive their signals not from other neurons but from \emph{receptors}.
The motor neurons transmits signals not to other neurons but to one or more \emph{effectors} igniting chemical reactions like the ones responsible for the contraction of muscles.
The sensory neurons collect information from the outside world which is passed on via the interneurons to various parts of the network. This defines the state of the system which is visible via a response in the motor neurons.
The ``end-station'' of an input is not necessarily a specific motor neuron. The inputs are collectively setting the whole network in different states, and thus produces different responses to different inputs.

\begin{figure}[htb]
\begin{center}
\includegraphics[width=0.95\columnwidth]{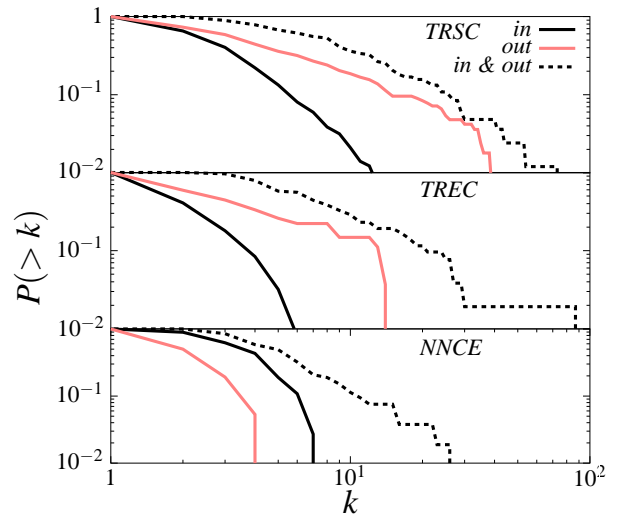}
\caption{The cumulative degree distribution of vertices with; only incoming arcs, only outgoing arcs, and with both incoming and outgoing arcs.}
\label{Fig2}
\end{center}
\end{figure}

In figure \ref{Fig2} the degree distribution of the different networks are plotted and one can observe that the transcriptional regulation networks (TRSC and TREC) are somewhat similar in the sense that the degree of the vertices with only incoming arcs have a lower degree than the rest of the vertices. 
This indicates that the proteins with no control and with a position in analogy of a laborer tend to be controlled by a few proteins and often just one protein which often has a high degree \cite{maslov:pro} and with mostly just outgoing arcs, a global controller.
In the neural network the situation is precisely the opposite, the sensors have in general very few outgoing arcs, in fact often just one.
The sensory neurons are in most cases connected to an interneuron of a relative high degree of incoming links from which it collects information from a number of sensory neurons. In only a few cases the sensory neurons are connected directly to a motor neuron. 

\section{Robustness via Structure}

Many networks are believed to have a modular structure with functional modules in which communication is more present than between the modules. In addition to just having separate functions this also minimizes the influence of a random change of the network. 
The modularity has been studied and detected in undirected networks \cite{bara:modhie,gir:alg,castel:comm} and network models \cite{my:sec}. In nature there are many things found or believed to be modular \cite{Lel:modcell,csete:eng} where the separate modules are responsible for different functions and together serve as a unit in a larger system. The modularity of the transcriptional regulation network of \emph{S.\ cerevisiae} (TRSC) has been studied \cite{ihm:yea}, and also the robustness in \cite{born:gen}.
Some modules are more important than others, and by removing a unit more or less of the function of the total system is removed.
Since the transcriptional regulation networks serves as regulating systems of the production of various proteins with different tasks they need to be constructed to remain most of the functions even if subjected to random removal or random changes of proteins. Random changes are naturally present via mutations in the \emph{DNA}.
Besides the fact that the \emph{DNA} contains ``garbage'' which reduces the probability of removing important functions, one could ask if the transcriptional regulation networks have evolved to a structure which is robust to mutations and if it is possible to reveal and quantify the robustness with some measure of the structure?

In the literature there are a number of different measures of prestige and influence, see eg. \cite{lin:soc}. Let $D_i$ be the number of downstream vertices of a vertex $v_i$, and define the influence $I_i$ of the vertex $v_i$ to be the fraction of vertices in the network which is downstream of vertex $v_i$,

\begin{equation}
I_i = \frac{D_i}{N-1}
\end{equation}

The distribution of the influence $P(I)$ of the vertices in the network gives information of how the influence and control are distributed in the network.
Moreover, the distribution $P(I)$ also gives information of how large fraction of the network that maximally (and typically) can be affected by a random change.
For the function of the network to be stable to changes, the structure has to be designed in a way where most vertices only influence a low fraction of the vertices in the network. But even though the stability is important, the function of the network might anyway need some vertices of great influence or control, global controller proteins. The global controllers are needed for the response to nutritional elements C, O, N, P, heat shock, growth rate, and more.

\begin{figure}[htb]
\begin{center}
\includegraphics[width=0.95\columnwidth]{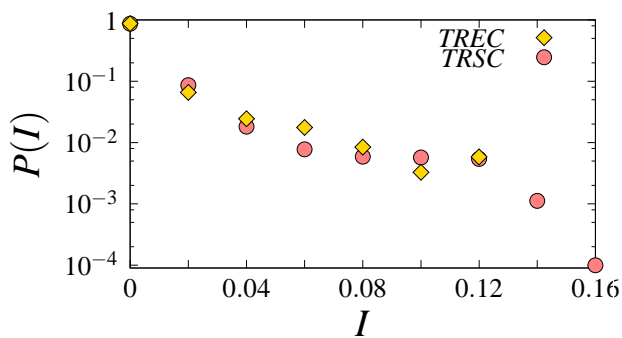}
\caption{The distribution of influence $P(I)$ of TREC and TRSC. The data is binned.}
\label{Fig3}
\end{center}
\end{figure}

In figure \ref{Fig3} the distribution $P(I)$ of TRSC and TREC is plotted.
From the plot one can see that in TREC and TRSC that most of the proteins control only a very little part of the network.
The implication of this is that a random mutation or deletion of a vertex (protein) affects only a very small part of the network.
In the case of NNCE, the situation is the reverse as can be seen in table \ref{Tab1}.

\begin{table}[htb]
\begin{center}
\begin{tabular*}{0.75\columnwidth}{l@{\extracolsep{\fill}}ll} \hline
$I$  & $P(I)$ & $\Delta S_f$  \\ \hline \hline
0.0  & 0.075  & 0.0 \\
0.007  & 0.014  & 0.0030(0) \\
0.939  & 0.854  & 0.0052(6) \\
0.942  & 0.043  & 0.0035(8) \\
0.946  & 0.014  & 0.0051(7) \\ \hline
\end{tabular*}
\caption{The influence $I$, the fraction of neurons of influence $P(I)$, and the effect $\Delta S_f$ of removal of a neuron of influence $I$.}
\label{Tab1}
\end{center}
\end{table}

 A random removal or damage of a neuron can possibly affect the whole network. How and to which extent probably depends on the exact dynamics of the network and the situation. The network is still likely to be connected after a random vertex removal because of the high density of arcs, however because of the high influence of the vertices a random removal of a neuron will possibly change the state of a large fraction of the other neurons, and thus the response to different inputs/stimuli. This is analyzed in the next section.

\section{Robustness via Dynamics}

To analyze whether the influence $I$ of a vertex is of importance when considering vertex removal, two simple models are used, where one captures the nature of the interactions of the trancriptional networks and the other the neural networks.
In \cite{kauf:bool}, the trancriptional network of \emph{S.\ cerevisiae} (TRSC) is analyzed in terms of boolean network models with the aim of determining feasible rule structures.
In their paper they find that many of the generated networks are shown to have a substantial part which is frozen in the sense that the final state is the same regardless of the initial states.
As described before the vertices in the trancriptional networks consists of proteins and the arcs represents one protein's regulation of another, in which the regulation can be either a positive regulation, an activator protein, or a negative regulation, a repressor.
Also a study of the robustness of transcriptional regulation networks with the use of neural networks are done in \cite{kim:top}.

\begin{figure}[htb]
\begin{center}
\includegraphics[width=0.95\columnwidth]{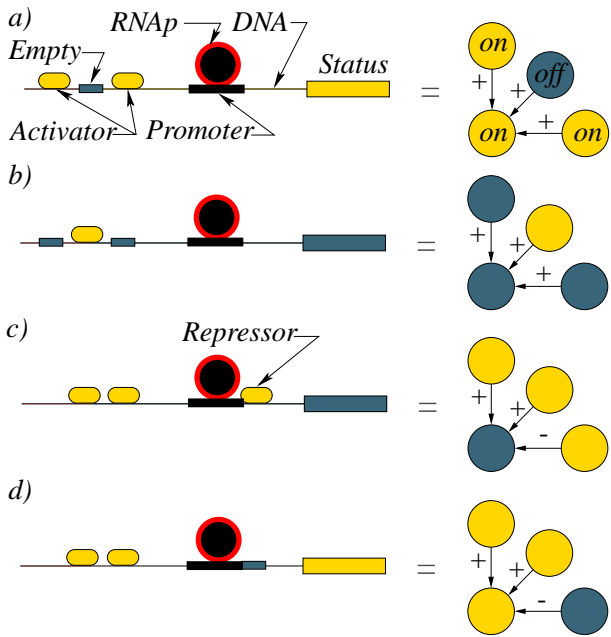}
\caption{The update rules for the protein regulation networks. The positively regulating proteins, activator proteins, are treated with a majority rule illustrated in a) and b) and is override by a negative regulation, repressor, which is illustrated in c) and d).}
\label{Fig4}
\end{center}
\end{figure}

In the model that we use to simulate the transcriptional regulation, the state of a gene coding for a specific protein $v_i$ has two values, expressed or not expressed; {\it on} or {\it off}. 
If the state of the gene coding for a specific protein is {\it off} there is no production, or at most a very small production, of the protein and is therefore not considered to be present in the system.
If the state is {\it on} there is a production enough for the protein considered to be present in the system.
The state of a gene coding for a protein $v_i$ is determined and regulated by the proteins $v_j$ with arcs pointing towards $v_i$. The proteins with no incoming arcs are determined from the initiation and can be considered as different environmental settings. The rules for how the update is done can be summarized as:
\begin{itemize}
\item All vertices are randomly initiated with the value {\it on} or {\it off}.
\item The vertices $v_i$ are then updated sequentially with the following rule until a final state is achieved:
  \begin{itemize}
  \item The state of all vertices $v_j$ pointing at $v_i$ are determined.
  \item If the state of a protein $v_j$ with negative regulation (repressor) is {\it on}, the state of protein $v_i$ is {\it off}.
  \item If no negative regulation is present, the regulation follows a majority rule and the state of the protein $v_i$ is {\it on/off} if the {\it majority} of the state of the positive regulating proteins $v_j$ are {\it on/off}.
  \end{itemize}
\end{itemize}
The update is illustrated in figure \ref{Fig4}. The motivation for the model follows from the nature of the interactions. Negative regulation by a repressor blocks the production of a protein by binding to an operator downstream of the promoter of the gene that codes to the specific protein. 
When the repressor sits downstream of the promoter it stops the transcription of the downstream gene from RNAp.
Positive regulation by an activator enhances the probability of RNAp to attach to the promoter of the gene and thereby the transcription of the gene that codes for the specific protein.

\begin{figure}[htb]
\begin{center}
\includegraphics[width=0.95\columnwidth]{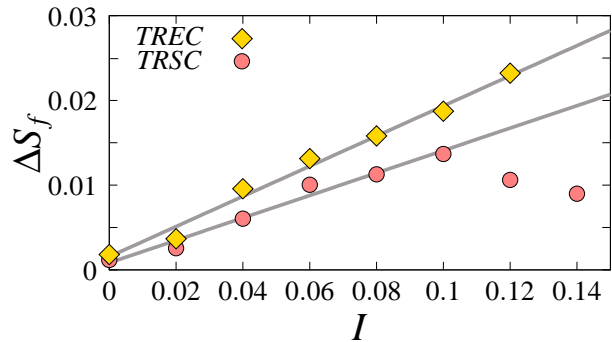}
\caption{The difference of the final state $\Delta S_f$ versus vertex removal of a vertex with influence $I$. The lines are straight line approximations to guide the eye and the plotted data is binned. The errors are smaller than the symbol sizes.}
\label{Fig5}
\end{center}
\end{figure}

The state of the total system is represented by a vector of dimension $N$, $(S_1,S_2,...,S_N)$. To investigate the effect of a random vertex deletion and to which extent the influence of the vertex plays a role for the state of the system, the final state obtained from an initial configuration is compared with the corresponding final state from the same initial configuration but with a vertex deleted from the network.
The relationship of the influence of the removed vertex and the effect of the influence of the final state is demonstrated in figure \ref{Fig5}.
$\Delta S_f$ is the fraction of vertices (proteins) in the network having a different final state after the removal of a vertex $v$.
Only the initial configurations that converge to a final state is considered.
The fraction of initial states that converge to a final state is approximately $1$ for TREC and $0.7$ for TRSC. 
The initial configurations that do not converge to a final state ends up in an oscillatory state, and they are not considered or investigated here.
As one can see the overall behavior is that the difference in the final state $\Delta S_f$ has a somewhat linear behavior of the influence $I$ of the vertex $v$ being removed.
The plot of TRSC do not follow the straight line approximation for larger values of $I$, which might indicate that the measure of influence used here is not perfectly suited for the applied dynamics.

Since the networks have a fix structure, the values for the difference in the final states $\Delta S_f$ for the different influences $I$ does not all follow the approximated straight line as can be observed, but the calculated error for the individual values of $\Delta S_f$ are nevertheless small.
Since the dynamics incorporates a majority rule, the effect of a vertex removal decays with the distance from the removed vertex and therefore only a fraction of the downstream vertices get a different state after the removal.
How large fraction depends therefore on the structure of the network and the typical distance to the downstream vertices from the removed one.

\begin{figure}[htb]
\begin{center}
\includegraphics[width=0.95\columnwidth]{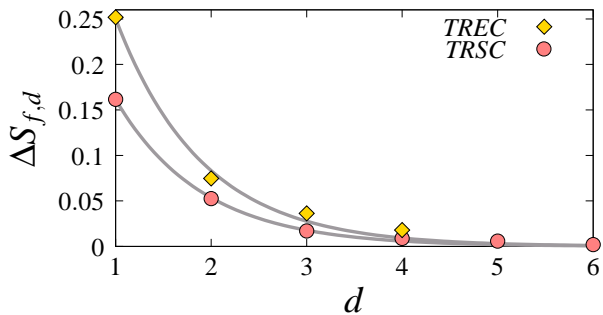}
\caption{The fraction of vertices with a different state $\Delta S_{f,d}$ plotted against distance of the removed vertex. The data is fitted to exponential functions. The errors are smaller than the symbol sizes.}
\label{Fig6}
\end{center}
\end{figure}

Figure \ref{Fig6} shows the fraction of all vertices with distance $d$ from the removed vertex which have a different final state $\Delta S_{f,d}$ compared with the final state before the removal.
Except from the exponential decay, one can also observe that the longest directed path is only of four steps in TREC and of six in TRSC.
As a comparison the diameter in TREC is $13$ and in TRSC $14$.
Since the fraction of vertices with a different state drops exponentially with distance a refinement of the measure of influence used here would be the measure of \emph{proximity prestige} (see \cite{lin:soc}), which is a vertex's number of downstream vertices normalized with their average distance to the vertex.

NNCE is applied to a similar dynamics as the transcriptional regulation networks, with the only difference that there are no negative regulations which overrule the positive ones.
The neurons are thus treated as \emph{McCullough-Pitts} neurons \cite{pitts:neur,hertz} with binary states {\it on/off}, and with equal and positive synaptic coupling strength (excitatory) and with a threshold of $n_i/0.5$, where $n_i$ are the number of inputs.
All neurons are considered to be excitatory, that is in a state {\it off} when the input is below the threshold and {\it on} if the input is above the threshold.
The update of the state of each neuron is therefore simply a majority rule, that is the state $S_i$ of a neuron $v_i$ is {\it on/off} if the majority of the state of the incoming signals are {\it on/off}.
If there are no majority, that is the number of {\it on} inputs are equal to the number of {\it off} inputs the state of the neuron defined to be {\it on}.
Since the influence of the neurons in the network are concentrated to a value around $I=0.94$ a linear dependence of the difference in the state $\Delta S_f$ to the influence $I$ is thus not achievable.
However, we can still get information of the effect of a vertex removal just by looking at the average difference in the final state from a deletion of a neuron. As before the results are averaged by a number of different initial configurations and different vertex removals.

\begin{figure}[htb]
\begin{center}
\includegraphics[width=0.95\columnwidth]{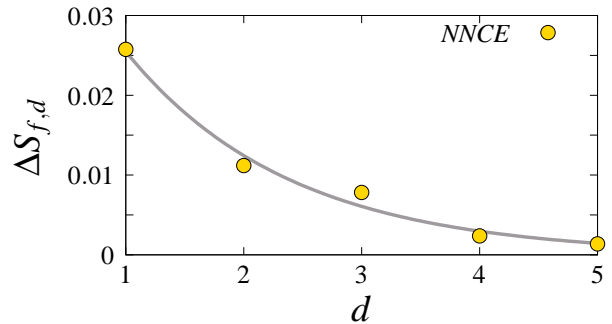}
\caption{The fraction of vertices with a different state $\Delta S_{f,d}$ plotted against distance of the removed vertex. The data is fitted to an exponential function.}
\label{Fig7}
\end{center}
\end{figure}

The results of the simulations done for NNCE are summarized in table \ref{Tab1}.
As one can see, the change in the state of the system is very small even if the influence of the removed vertex is large. The fact that the network is very dense and that the dynamics follow a majority rule implies that the change in the state of the system from a deletion of an individual neuron is small.
The change in the state from the removal of a single neuron is simply averaged out in most cases, but there are of course changes in the state of some neurons located nearby the removed neuron.
A further study would be to see how the state of the system responds when deleting a neighborhood of neurons to resemble a more realistic physical damage.
One of the conclusions one can make is that even if the influence of most vertices are large, the dynamics put on the network results in a situation where the network is not very affected by a random removal of a single neuron.
Figure \ref{Fig7} shows the decay of the fraction of changed states with the distance, and like the transcriptional regulation networks the decay fits well to an exponential decay.

\section{Summary}
The two directed types of networks analyzed here are shown to have a different structure in various measures which incorporates the direction of the edges.
The neural network of \emph{C.\ elegance} consists of mostly vertices with both incoming and outgoing arcs, interneurons, possibly due to the fact that it is an information network in which information is processed and spread between different parts of the network. 
There is also a fraction of vertices with only outgoing arcs, sensory neurons, that feed the network with external information. Finally, there is a small fraction of vertices with only incoming arcs, motor neurons, responsible for igniting chemical reactions like contraction of muscles.

We find that the protein regulation networks of \emph{E.\ coli} and \emph{S.\ cerevisae} both show a small fraction of proteins with both incoming and outgoing arcs, and the dominating part of the network consists of proteins that only have incoming arcs, which we term laborers in the analogy of a human laborer which only has a small influence in the system he or she works in.
The laborers are possibly used as building blocks or as components in different biochemical processes.
There are also proteins that have a large number of outgoing arcs and thus globally controls the production of many proteins, where most of them are laborers.

The influence of the vertices, defined for a vertex as the fraction of vertices in the network situated downstream of the vertex, are limited for most vertices in the transcriptional regulation networks. The neural network is showing a substantial part of the network to influence almost all vertices in the network.
We simulate the biological processes on the networks and we investigate relationship of the influence of a removed vertex with the change in the state of the biological system.
In the protein regulation networks the effect of a random removal are limited in most cases due to the fact that the influence of most proteins are restricted to a small number of proteins. However a removal of a protein of great influence changes the state of the system more since the change in the state is shown to increase, with some exceptions, linearly with the influence. 
In the case of the neural network where the influence of most vertices are fairly large, the great number of arcs suppresses the effect of a removal of an individual neuron since the state of each neuron obeys a majority rule of the states of the incoming signals.
Since the changes in the states of the vertices due to a vertex removal is shown to decay exponentially with the distance from the removed vertex, a proper refinement of the measure of influence could be to include the typical distance to the downstream vertices in the measure of influence, just like the measure of \emph{proximity prestige}.
A remark with the previous observations in mind would therefore be that the influence or prestige in a network (directed or not) probably to a large extent depends on the dynamics applied to the network, and therefore every investigation of prestige or influence should be in conjunction with the dynamics applied to the network.

Many thanks to Petter Holme and Martin Rosvall for your helpful comments and beneficial discussions contributing to this paper.

\bibliographystyle{apsrev}
\bibliography{mybib}

\begin{thebibliography}{26}
\expandafter\ifx\csname natexlab\endcsname\relax\def\natexlab#1{#1}\fi
\expandafter\ifx\csname bibnamefont\endcsname\relax
  \def\bibnamefont#1{#1}\fi
\expandafter\ifx\csname bibfnamefont\endcsname\relax
  \def\bibfnamefont#1{#1}\fi
\expandafter\ifx\csname citenamefont\endcsname\relax
  \def\citenamefont#1{#1}\fi
\expandafter\ifx\csname url\endcsname\relax
  \def\url#1{\texttt{#1}}\fi
\expandafter\ifx\csname urlprefix\endcsname\relax\def\urlprefix{URL }\fi
\providecommand{\bibinfo}[2]{#2}
\providecommand{\eprint}[2][]{\url{#2}}

\bibitem[{\citenamefont{Barab\'{a}si and Albert}(1999)}]{ba:model}
\bibinfo{author}{\bibfnamefont{A.-L.} \bibnamefont{Barab\'{a}si}}
  \bibnamefont{and} \bibinfo{author}{\bibfnamefont{R.}~\bibnamefont{Albert}},
  \bibinfo{journal}{Science} \textbf{\bibinfo{volume}{286}},
  \bibinfo{pages}{509} (\bibinfo{year}{1999}).

\bibitem[{\citenamefont{Newman}(2002)}]{mejn:assmix}
\bibinfo{author}{\bibfnamefont{M.~E.~J.} \bibnamefont{Newman}},
  \bibinfo{journal}{Phys. Rev. Lett.} \textbf{\bibinfo{volume}{89}},
  \bibinfo{pages}{208701} (\bibinfo{year}{2002}).

\bibitem[{\citenamefont{Freeman}(1977)}]{free:cen}
\bibinfo{author}{\bibfnamefont{L.}~\bibnamefont{Freeman}},
  \bibinfo{journal}{Sociometry} \textbf{\bibinfo{volume}{40}},
  \bibinfo{pages}{35} (\bibinfo{year}{1977}).

\bibitem[{\citenamefont{Barth\'{e}lemy}()}]{bart:betw}
\bibinfo{author}{\bibfnamefont{M.}~\bibnamefont{Barth\'{e}lemy}},
  \bibinfo{note}{e-print cond-mat/0309436}.

\bibitem[{\citenamefont{Newman}()}]{mejn:rwbetw}
\bibinfo{author}{\bibfnamefont{M.~E.~J.} \bibnamefont{Newman}},
  \bibinfo{note}{e-print cond-mat/0309045}.

\bibitem[{\citenamefont{Newman}(2003)}]{mejn:rev}
\bibinfo{author}{\bibfnamefont{M.~E.~J.} \bibnamefont{Newman}},
  \bibinfo{journal}{SIAM Rev.} \textbf{\bibinfo{volume}{45}},
  \bibinfo{pages}{167} (\bibinfo{year}{2003}).

\bibitem[{\citenamefont{Dorogovtsev and Mendes}(2002)}]{doromen:rev}
\bibinfo{author}{\bibfnamefont{S.~N.} \bibnamefont{Dorogovtsev}}
  \bibnamefont{and} \bibinfo{author}{\bibfnamefont{J.~F.~F.}
  \bibnamefont{Mendes}}, \bibinfo{journal}{Adv. Phys.}
  \textbf{\bibinfo{volume}{51}}, \bibinfo{pages}{1079} (\bibinfo{year}{2002}).

\bibitem[{\citenamefont{Bornholdt and Schuster}(2002)}]{born:book}
\bibinfo{author}{\bibfnamefont{S.}~\bibnamefont{Bornholdt}} \bibnamefont{and}
  \bibinfo{author}{\bibfnamefont{H.~G.} \bibnamefont{Schuster}},
  \emph{\bibinfo{title}{Handbook of Graphs and Networks - From the Genome to
  the Internet}} (\bibinfo{publisher}{Wiley-VCH}, \bibinfo{address}{Berlin},
  \bibinfo{year}{2002}).

\bibitem[{\citenamefont{White et~al.}(1986)\citenamefont{White, Southgate,
  Thompson, and Brenner}}]{white:420}
\bibinfo{author}{\bibfnamefont{J.~G.} \bibnamefont{White}},
  \bibinfo{author}{\bibfnamefont{E.}~\bibnamefont{Southgate}},
  \bibinfo{author}{\bibfnamefont{J.~N.} \bibnamefont{Thompson}},
  \bibnamefont{and} \bibinfo{author}{\bibfnamefont{S.}~\bibnamefont{Brenner}},
  \bibinfo{journal}{Philos. Trans. Roy. Soc. Lond.}
  \textbf{\bibinfo{volume}{314}}, \bibinfo{pages}{1} (\bibinfo{year}{1986}).

\bibitem[{\citenamefont{Shen-Orr et~al.}(2002)\citenamefont{Shen-Orr, Milo,
  Mangan, and Alon}}]{alon:ecoli}
\bibinfo{author}{\bibfnamefont{S.}~\bibnamefont{Shen-Orr}},
  \bibinfo{author}{\bibfnamefont{R.}~\bibnamefont{Milo}},
  \bibinfo{author}{\bibfnamefont{S.}~\bibnamefont{Mangan}}, \bibnamefont{and}
  \bibinfo{author}{\bibfnamefont{U.}~\bibnamefont{Alon}},
  \bibinfo{journal}{Nature Genetics} \textbf{\bibinfo{volume}{31}},
  \bibinfo{pages}{64} (\bibinfo{year}{2002}).

\bibitem[{\citenamefont{Lee et~al.}(2002)\citenamefont{Lee, Rinaldi, Robert,
  Odom, Bar-Joseph, Gerber, Hannett, Harbison, Thompson, Simon
  et~al.}}]{Lee:yeast}
\bibinfo{author}{\bibfnamefont{T.~I.} \bibnamefont{Lee}},
  \bibinfo{author}{\bibfnamefont{N.~J.} \bibnamefont{Rinaldi}},
  \bibinfo{author}{\bibfnamefont{F.}~\bibnamefont{Robert}},
  \bibinfo{author}{\bibfnamefont{D.~T.} \bibnamefont{Odom}},
  \bibinfo{author}{\bibfnamefont{Z.}~\bibnamefont{Bar-Joseph}},
  \bibinfo{author}{\bibfnamefont{G.~K.} \bibnamefont{Gerber}},
  \bibinfo{author}{\bibfnamefont{N.~M.} \bibnamefont{Hannett}},
  \bibinfo{author}{\bibfnamefont{C.~T.} \bibnamefont{Harbison}},
  \bibinfo{author}{\bibfnamefont{C.~M.} \bibnamefont{Thompson}},
  \bibinfo{author}{\bibfnamefont{I.}~\bibnamefont{Simon}},
  \bibnamefont{et~al.}, \bibinfo{journal}{Science}
  \textbf{\bibinfo{volume}{298}}, \bibinfo{pages}{799} (\bibinfo{year}{2002}).

\bibitem[{\citenamefont{Purves et~al.}(2004)\citenamefont{Purves, Augustine,
  Fitzpatrick, Hall, LaMantia, McNamara, and Williams}}]{dale:neur}
\bibinfo{editor}{\bibfnamefont{D.}~\bibnamefont{Purves}},
  \bibinfo{editor}{\bibfnamefont{G.~J.} \bibnamefont{Augustine}},
  \bibinfo{editor}{\bibfnamefont{D.}~\bibnamefont{Fitzpatrick}},
  \bibinfo{editor}{\bibfnamefont{W.~C.} \bibnamefont{Hall}},
  \bibinfo{editor}{\bibfnamefont{A.-S.} \bibnamefont{LaMantia}},
  \bibinfo{editor}{\bibfnamefont{J.~O.} \bibnamefont{McNamara}},
  \bibnamefont{and} \bibinfo{editor}{\bibfnamefont{S.~M.}
  \bibnamefont{Williams}}, eds., \emph{\bibinfo{title}{Neuroscience}}
  (\bibinfo{publisher}{Sinauer Associates Incorporated}, \bibinfo{year}{2004}),
  \bibinfo{edition}{3rd} ed.

\bibitem[{\citenamefont{Maslov and Sneppen}(2002)}]{maslov:pro}
\bibinfo{author}{\bibfnamefont{S.}~\bibnamefont{Maslov}} \bibnamefont{and}
  \bibinfo{author}{\bibfnamefont{K.}~\bibnamefont{Sneppen}},
  \bibinfo{journal}{Science} \textbf{\bibinfo{volume}{296}},
  \bibinfo{pages}{910} (\bibinfo{year}{2002}).

\bibitem[{\citenamefont{Ravasz et~al.}(2002)\citenamefont{Ravasz, Somera,
  Mongru, Oltvai, and Barab\'{a}si}}]{bara:modhie}
\bibinfo{author}{\bibfnamefont{E.}~\bibnamefont{Ravasz}},
  \bibinfo{author}{\bibfnamefont{A.~L.} \bibnamefont{Somera}},
  \bibinfo{author}{\bibfnamefont{D.~A.} \bibnamefont{Mongru}},
  \bibinfo{author}{\bibfnamefont{Z.~N.} \bibnamefont{Oltvai}},
  \bibnamefont{and} \bibinfo{author}{\bibfnamefont{A.-L.}
  \bibnamefont{Barab\'{a}si}}, \bibinfo{journal}{Science}
  \textbf{\bibinfo{volume}{297}}, \bibinfo{pages}{1553} (\bibinfo{year}{2002}).

\bibitem[{\citenamefont{Girvan and Newman}(2002)}]{gir:alg}
\bibinfo{author}{\bibfnamefont{M.}~\bibnamefont{Girvan}} \bibnamefont{and}
  \bibinfo{author}{\bibfnamefont{M.~E.~J.} \bibnamefont{Newman}},
  \bibinfo{journal}{Proc. Natl. Acad. Sci. USA} \textbf{\bibinfo{volume}{99}},
  \bibinfo{pages}{7821} (\bibinfo{year}{2002}).

\bibitem[{\citenamefont{Radicchi et~al.}()\citenamefont{Radicchi, Castellano,
  Cecconi, Loreto, and Parisi}}]{castel:comm}
\bibinfo{author}{\bibfnamefont{F.}~\bibnamefont{Radicchi}},
  \bibinfo{author}{\bibfnamefont{C.}~\bibnamefont{Castellano}},
  \bibinfo{author}{\bibfnamefont{F.}~\bibnamefont{Cecconi}},
  \bibinfo{author}{\bibfnamefont{V.}~\bibnamefont{Loreto}}, \bibnamefont{and}
  \bibinfo{author}{\bibfnamefont{D.}~\bibnamefont{Parisi}},
  \bibinfo{note}{e-print cond-mat/0309488}.

\bibitem[{\citenamefont{Gr{\"o}nlund and Holme}(2004)}]{my:sec}
\bibinfo{author}{\bibfnamefont{A.}~\bibnamefont{Gr{\"o}nlund}}
  \bibnamefont{and} \bibinfo{author}{\bibfnamefont{P.}~\bibnamefont{Holme}},
  \bibinfo{journal}{Phys. Rev. E.}  (\bibinfo{year}{2004}).

\bibitem[{\citenamefont{Hartwell et~al.}(1999)\citenamefont{Hartwell, Hopfield,
  Leibler, and Murray}}]{Lel:modcell}
\bibinfo{author}{\bibfnamefont{L.~H.} \bibnamefont{Hartwell}},
  \bibinfo{author}{\bibfnamefont{J.~J.} \bibnamefont{Hopfield}},
  \bibinfo{author}{\bibfnamefont{S.}~\bibnamefont{Leibler}}, \bibnamefont{and}
  \bibinfo{author}{\bibfnamefont{A.~W.} \bibnamefont{Murray}},
  \bibinfo{journal}{Nature} \textbf{\bibinfo{volume}{402}},
  \bibinfo{pages}{C47} (\bibinfo{year}{1999}).

\bibitem[{\citenamefont{Csete and Doyle}(2002)}]{csete:eng}
\bibinfo{author}{\bibfnamefont{M.~E.} \bibnamefont{Csete}} \bibnamefont{and}
  \bibinfo{author}{\bibfnamefont{J.~C.} \bibnamefont{Doyle}},
  \bibinfo{journal}{Science} \textbf{\bibinfo{volume}{295}},
  \bibinfo{pages}{1664} (\bibinfo{year}{2002}).

\bibitem[{\citenamefont{Ihmels et~al.}(2002)\citenamefont{Ihmels, Friedlander,
  Bergmann, Sarig, Ziv, and Barkai}}]{ihm:yea}
\bibinfo{author}{\bibfnamefont{J.}~\bibnamefont{Ihmels}},
  \bibinfo{author}{\bibfnamefont{G.}~\bibnamefont{Friedlander}},
  \bibinfo{author}{\bibfnamefont{S.}~\bibnamefont{Bergmann}},
  \bibinfo{author}{\bibfnamefont{O.}~\bibnamefont{Sarig}},
  \bibinfo{author}{\bibfnamefont{Y.}~\bibnamefont{Ziv}}, \bibnamefont{and}
  \bibinfo{author}{\bibfnamefont{N.}~\bibnamefont{Barkai}},
  \bibinfo{journal}{Nature Genetics} \textbf{\bibinfo{volume}{31}},
  \bibinfo{pages}{370} (\bibinfo{year}{2002}).

\bibitem[{\citenamefont{Bornholdt}(2001)}]{born:gen}
\bibinfo{author}{\bibfnamefont{S.}~\bibnamefont{Bornholdt}},
  \bibinfo{journal}{Biological Chemistry} \textbf{\bibinfo{volume}{382}},
  \bibinfo{pages}{1289} (\bibinfo{year}{2001}).

\bibitem[{\citenamefont{Lin}(1976)}]{lin:soc}
\bibinfo{author}{\bibfnamefont{N.}~\bibnamefont{Lin}},
  \emph{\bibinfo{title}{Foundations of Social Research}}
  (\bibinfo{publisher}{McGraw-Hill}, \bibinfo{address}{New York},
  \bibinfo{year}{1976}).

\bibitem[{\citenamefont{Stuart et~al.}(2003)\citenamefont{Stuart, Carsten,
  Bj\"{o}rn, and Carl}}]{kauf:bool}
\bibinfo{author}{\bibfnamefont{K.}~\bibnamefont{Stuart}},
  \bibinfo{author}{\bibfnamefont{P.}~\bibnamefont{Carsten}},
  \bibinfo{author}{\bibfnamefont{S.}~\bibnamefont{Bj\"{o}rn}},
  \bibnamefont{and} \bibinfo{author}{\bibfnamefont{T.}~\bibnamefont{Carl}},
  \bibinfo{journal}{Proc. Natl. Acad. Sci. USA} \textbf{\bibinfo{volume}{100}},
  \bibinfo{pages}{14796} (\bibinfo{year}{2003}).

\bibitem[{\citenamefont{S and JN}(2003)}]{kim:top}
\bibinfo{author}{\bibfnamefont{K.}~\bibnamefont{S}} \bibnamefont{and}
  \bibinfo{author}{\bibfnamefont{W.}~\bibnamefont{JN}}, in
  \emph{\bibinfo{booktitle}{Systems, Man and Cybernetics, 2003 IEEE
  International Conference}} (\bibinfo{address}{Washington, D.C., USA},
  \bibinfo{year}{2003}), \bibinfo{number}{4}, pp. \bibinfo{pages}{3969--3975}.

\bibitem[{\citenamefont{McCulloch and Pitts}(1943)}]{pitts:neur}
\bibinfo{author}{\bibfnamefont{W.~S.} \bibnamefont{McCulloch}}
  \bibnamefont{and} \bibinfo{author}{\bibfnamefont{W.}~\bibnamefont{Pitts}},
  \bibinfo{journal}{Bulletin of Mathematical Biophysics}
  \textbf{\bibinfo{volume}{5}}, \bibinfo{pages}{115} (\bibinfo{year}{1943}).

\bibitem[{\citenamefont{Hertz et~al.}(1991)\citenamefont{Hertz, Krogh, and
  Palmer}}]{hertz}
\bibinfo{author}{\bibfnamefont{J.}~\bibnamefont{Hertz}},
  \bibinfo{author}{\bibfnamefont{A.}~\bibnamefont{Krogh}}, \bibnamefont{and}
  \bibinfo{author}{\bibfnamefont{R.~G.} \bibnamefont{Palmer}},
  \emph{\bibinfo{title}{Introduction to the theory of neural computation}}
  (\bibinfo{publisher}{Addison-Wesley}, \bibinfo{address}{Redwood City},
  \bibinfo{year}{1991}).

\end{thebibliography}

\end{document}